\documentclass[showpacs,prl,twocolumn,floatfix,reprint]{revtex4}
\usepackage{graphicx,amssymb,amsmath,xcolor}
\usepackage{stmaryrd}

\begin{document}

\title{Towards a classification of planar maps} 

\author{Alexandre Diet}
\affiliation{Institut de Physique Th\'{e}orique, CEA, CNRS-URA 2306, F-91191, 
Gif-sur-Yvette, France}

\author{Marc Barthelemy}
\email{marc.barthelemy@ipht.fr}
\affiliation{Institut de Physique Th\'{e}orique, CEA, CNRS-URA 2306, F-91191, 
Gif-sur-Yvette, France}
\altaffiliation{CAMS (CNRS/EHESS) 190-198, avenue de France, 75244 Paris Cedex 13, France}

\begin{abstract}

Planar graphs and their spatial embedding -- planar maps -- are used in many different fields due to their ubiquity in the real world (leaf veins in biology, street patterns in urban studies, etc.) and are also fundamental objects in mathematics and combinatorics. These graphs have been well described in the literature, but we do not have so far a clear way to cluster them in different families. A typology of planar maps would be very useful and would allow to monitor their changes, to compare them with each other, or to correlate their structure with other properties. Using an algorithm which merges recursively the smallest areas in the graph with the largest ones, we plot the Gini coefficient of areas of cells and obtain a profile associated to each network. We test the relevance of these `Gini profiles' on simulated networks and on real street networks of Barcelona (Spain), New York City (USA), Tokyo (Japan), and discuss their main properties. We also apply this method to the case of Paris (France) at different dates which allows us to follow the structural changes of this system. Finally, we discuss the important ingredient of spatial heterogeneity of real-world planar graphs and test some ideas on Manhattan and Tokyo. Our results show that the Gini profile encodes various informations about the structure of the corresponding planar map and represents a good candidate for constructing relevant classes of these objects.

\end{abstract}


\maketitle

\section{Introduction}

A graph is planar if it can be drawn on the plane in such a way that no edges cross each other. Such graphs appear in many fields: mathematics and combinatorics \cite{Tutte1963,Bouttier:2004}, theoretical physics \cite{Ambjorn1997}, biology \cite{Mileyko2012} or urban studies \cite{Barthelemy2011}. In real-world applications, graphs exists under the form of an embedding in the 2d physical world and these objects are usually called planar maps (we note here that the embedding is not necessarily planar even if the graph is itself planar \cite{Barthelemy:2018}). A particularly important application concerns transportation networks such as streets and roads. In this case, nodes are intersections and the links are the segments of roads between nodes. Most of these networks can be considered as planar as they must intersect at a crossroad. Obviously we have bridges and tunnels but there is usually some interchange allowing to connect these roads on a relatively small scale so that the planarity defect is localized in space (a complete study of planarity for 50 cities worldwide can be found in \cite{Boeing:2018}). 

There are many attempts to characterize urban networks and planar graphs \cite{Haggett:1969, Xie:2007, Louf:2014b} (see also \cite{Barthelemy:2018} and references therein), and to classify them \cite{Katifori2012, Mileyko2012, Louf:2014b}, but it is fair to say that we still do not have a simple and robust method to achieve this goal. A classification of graphs could be useful in many fields. For cities, the street network is a simplified description of their structure and shape. It is therefore important to be able to characterize in a simple way this network, allowing for a comparison between cities, or to correlate the urban form with other quantities such as urban form, energy use, heat island effect \cite{Sob:2018}, etc. In addtion, the existence of different classes could point to simple mechanisms governing the evolution of cities. More generally, a good characterization should wash out all irrelevant details but should keep important  structural determinants. The balance between these two antagonistic requirements 
 is the key for a successful characterization of planar maps that can be used for
 a typology, and this paper discusses a candidate for such a tool.

In this paper, we will first (briefly) review the various attempts to classify planar graphs. In particular, we will detail the algorithm proposed simultaneously by Katifori and Magnasco \cite{Katifori2012} and Mileyko et al. \cite{Mileyko2012} for characterizing leaf veination patterns. lesf veins have various diameters and this algorithm was naturally developped for weighted networks. Here, we adapt this algorithm to the general case of non-weighted network and propose an indicator - the Gini profile - that describes the organization of faces (blocks in the case of street networks). Although our arguments are valid for all planar maps, we focus on the problem of characterizing street networks. We will apply our method to graphs generated \textit{in silico}, and we will discuss the application of this method to real-world cities such as Barcelona, Tokyo, New York city and Paris (for which we have the map for different dates). We end this paper by a discussion about perspectives for this method.

\section{The merging algorithm}

\begin{figure}
\begin{center}
\includegraphics[scale=0.5]{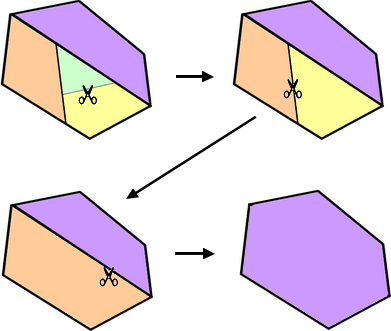}

\

\includegraphics[scale=0.7]{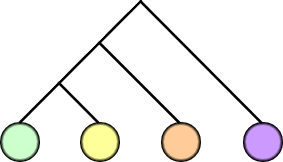}
\end{center}
\caption{Top: Visual representation of the merging algorithm \cite{Katifori2012,Mileyko2012}. The width of a line represents its diameter. Bottom: Resulting binarty tree that represents the merging process. The color of terminal nodes shows the corresponding cell.}
\label{algokatifori}
\end{figure}

In combinatorics, Bouttier, Di Francesco, and Guitter \cite{Bouttier:2004} discussed an exact mapping between a rooted planar map and a tree but this method seems difficult to use for constructing classes of graphs as the tree depends strongly on the root chosen. In \cite{Louf:2014b} a method for clustering planar maps was proposed and relied on the statistics of the shape and area size of cells. However, this method `deconstructs' in some way the network and we loose spatial correlations between blocks. We would like to keep track of these correlations and to find a simple way for characterizing the  spatial organization of the cells that is at the heart of the structure of planar maps. In order to make a first step towards this difficult task, we elaborate here on a method proposed previously \cite{Katifori2012,Mileyko2012} for weighted planar graphs. This method constructs an approximate mapping between a weighted planar map and a binary tree, and although it is not a bijection, it provides a simple and flexible framework.  In these studies, the weight corresponds to the diameter of veins in leaves. In the planar map, we thus have edges (veins here) that connect at nodes and form a set of faces that we will also call cells, or in the context of cities, blocks. The main steps in this algorithm are the following (see also Fig.~\ref{algokatifori}, top):

\begin{enumerate}
\item Associate to each edge $i$ its diameter $\delta_{i}$ (if there is a couple $i,j$ such that $\delta_{i} = \delta_{j}$, one can infinitesimally perturb $\delta_{i}$ or $\delta_{j}$ so that we can rank all the edges in a strict order). 

\item Find the smallest diameter $\delta_{m}$ and its associated edge $m$ (also called the `weakest edge').

\item Merge the two cells separated by the edge $m$. The new cell has an area equal to the sum of the merged cells, and keeps the same neighbors.

\item Go back to (2) until there are no more edges/until there exist only one cell.
\end{enumerate}

This algorithm allows to construct a binary tree where initial cells are the leaves (or terminal nodes) of this tree and the other nodes correspond to the fusion of different cells: two nodes are connected in this tree if the corresponding cells merge in the algorithm (see Fig.~\ref{algokatifori}, bottom). This algorithm thus maps a planar graph to a binary tree which can then be characterized with various measures (it is indeed in general easier on a tree). In \cite{Katifori2012}, the main tool used to characterize such trees is the asymmetry \cite{VanPelt1992}, a metric relative to their topological structure. 

However, in order to run the algorithm and build a binary tree for any planar map, the algorithm needs to be modified. Indeed, in a non-weighted planar map, we need a way to characterize the `weakness' of an edge for merging recursively the faces of the network. In order to setup this merging process we essentially need to choose the  cells that will be merged together. Many choices are possible: we can choose the smallest edge or with the smallest betweenness centrality, etc. but after various tests we decided to choose the area of cells as
the main indicator. Once the smallest cell is found we merge it to one of its neighbor and here also many choices are possible. We found that the most effective algorithm (for characterizing the networks) is to choose the neighbor with the largest area size. These choices can be motivated by the fact that in urban networks, we observe a hierarchical structure of blocks and the process here thus acts as a coarse-graining that eliminates the small scale details. With these choices, the merging procedure, which will be called `decimation' here is described by the following algorithm:

\begin{enumerate}
\item Associate to each cell $i$ its area $a_{i}$ (if there are two cells $i,j$ such that $a_{i} = a_{j}$, one can infinitesimally perturb $a_{i}$ or $a_{j}$).

\item Find the smallest area $a_{m}$ and its associate cell $m$.

\item Merge the cell $m$ with its neighbor of largest area size. The new cell has an area equal to the sum of the merged cells and keeps the same neighbors.

\item Go back to (2) until there is only one cell left.

\end{enumerate}

This algorithm will produce a sequence of cells whose sum is constant and equal to the total area size of the map. We now need a measure that can monitor the evolution of the graph during this process and that will reveal specific features of the planar map. In the next section we will discuss this and propose a characterization, the `Gini profile'.

\section{Characterizing the decimation process}

In order to devise and to test a characterization of the decimation process we will first
study three types of graphs (all constructed on a set of points), that can be similar but with intrinsic differences. These three types of graphs are: (i) the Poisson-Voronoi (PV), (ii) the Gabriel graph (GG) and (iii) the relative neighborhood graph (RN). The PV comes from the application of the Voronoi diagram \cite{Aurenhammer1991} algorithm to a set of $N$ points randomly chosen in the $[0,1] \times [0,1]$ square. The GG and RNG (see for example \cite{Aldous2010, Barthelemy:2018}) are subsets of the Delaunay graphs, and are also computed from $N$ points randomly selected in the $[0,1] \times [0,1]$ square. Specific realization of these graphs are shown in Fig.~\ref{extrees}.

\subsection{Decimation tree}

In \cite{Katifori2012}, various tools are used in order to characterize the decimation tree. These trees display particular structures (for example, what Katifori and Magnasco called `multiplicative' and `additive' trees), but for large graphs this characterization is difficult to use. 

\begin{figure*}[!ht]
\begin{tabular}{c|c|c}
Gabriel Graph	& Relative Neighborhood Graph	& Random Voronoi Graph	\\
\includegraphics[scale=0.3]{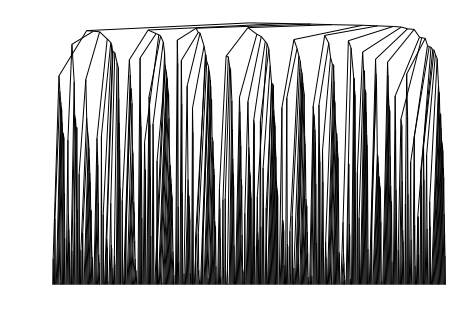}	
&\includegraphics[scale=0.3]{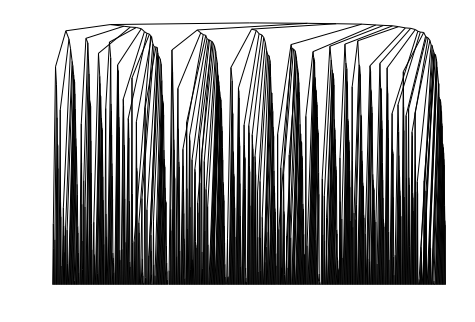} 
&\includegraphics[scale=0.3]{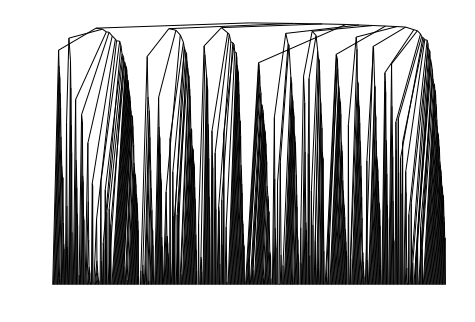}		\\
	\includegraphics[scale=0.3]{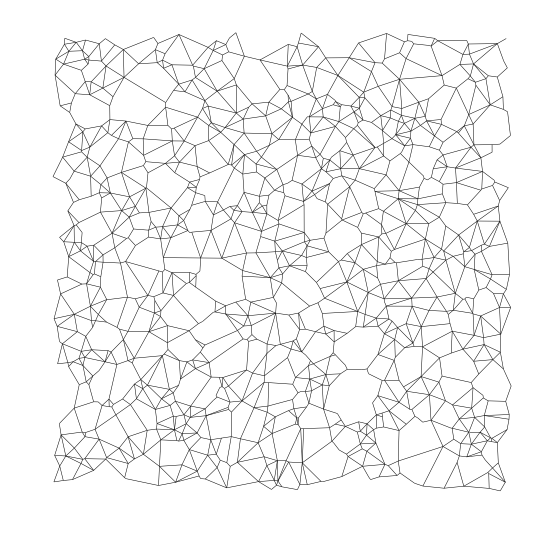} &
\includegraphics[scale=0.3]{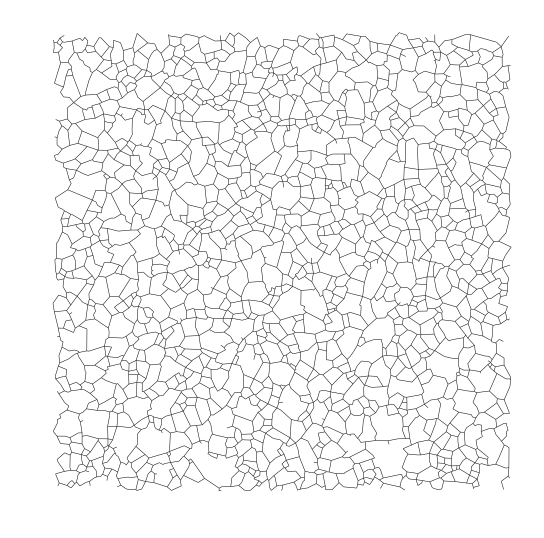} &
\includegraphics[scale=0.3]{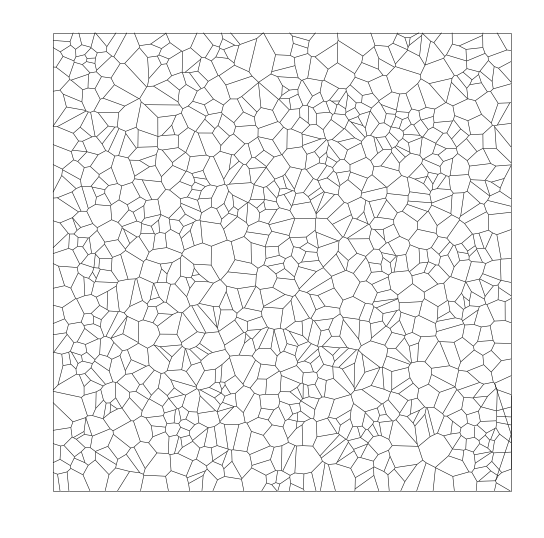}		\\
\end{tabular}
\caption{Decimation trees for different graphs. The Gabriel graph contains $924$ initial cells, the Relative Neighborhood one $1070$ initial cells and the Random Voronoi one $1000$ initial cells.}
\label{extrees}
\end{figure*}

For example, one can see that for about $1,000$ initial cells (see Fig.~\ref{extrees}), it is difficult to extract useful information from the decimation trees and to distinguish different types of graph. Furthermore, the asymmetry method used in \cite{Katifori2012} did not work as well on the graphs presented above. Indeed, the decimations for quite similar graphs (Gabriel, Relative Neighborhood, Random Voronoi) produce very similar results for the asymmetries making it difficult to distinguish these networks.

\subsection{Mean and variance of the areas distribution}

A simple idea is to consider the first two moments of the area size of cells during the decimation process. We will use the following notation:
\begin{itemize}
\item $N$ the initial number of cells in the graph

\item $t$ the step of the decimation, which will be called `time'. This quantities $t$ runs from $0$ to $N-1$, since there can only be $N-1$ fusions for $N$ initial cells.

\item $\tau=\dfrac{t}{N-1}$ is the normalized time, which will be useful to compare graphs with different $N$.

\item $(a_{1}, ..., a_{N})$ the area sizes of the initial cells.

\item $A=\sum\limits_{i=1}^{N} a_{i}$ the total area size constant in time.

\end{itemize}

\begin{figure}[!h]
\begin{center}
\includegraphics[scale=0.5]{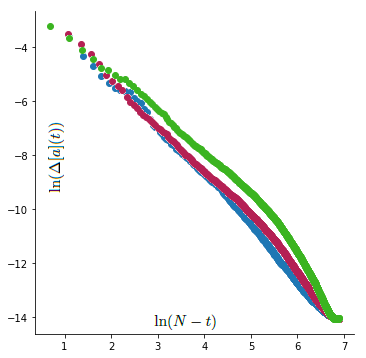}
\end{center}
\caption{Plot of $\ln(\Delta[a](t))$ vs. $\ln(N-t)$ for the Random Voronoi Graph (in green), the Relative Neighborhood Graph (in red) and the Gabriel Graph (in blue).}
\label{logvar}
\end{figure}

At time $t$, there are only $N-t$ cells left, and the area of the merged cells is the sum of the component cells. Thus, one sees that the mean area $\overline{a}(t)$ at time t is given by
\begin{align}
\overline{a}(t) = \frac{1}{N-t}\sum_ia_i=\frac{A}{N-t}
\end{align}
which shows that the average size is independent of any type of graph or decimation, and therefore cannot be used to characterize a graph. The next quantity is for example the variance $\Delta[a](t)$ of the area size and we studied this quantity numerically. Typical results for these different graphs are shown in Fig.~\ref{logvar}.
%
A power law fit of these curves suggest a behavior of the form $\Delta[a](t) \sim (1-\tau)^{-\alpha} \sim (N-t)^{-\alpha}$ with $\alpha\approx 2$ (these curves all display an average slope of $\alpha=2$ towards the end of the decimation and because of the change of axis, long times are on the left and short times are on the right). Here also, the variance seems to be independent from many details of the graph and can therefore not to be of any use for characterizing it. 

\subsection{The Gini profile}

The Gini coefficient \cite{Gini1921,Gini1936} first used in economics is a measure of inequality for a set of quantities. If $x_{i}$ represents the variable we want to study  (the income or the wealth of the person $i$, or the area size of a cell $i$), the Gini coefficient for a set of $n$ variables is then defined as follows \cite{Dixon:1987}
\begin{align}
G = \dfrac{\sum\limits_{i=1}^{n} \sum\limits_{j=1}^{n} \left| x_{i} - x_{j} \right|}{2 n \sum\limits_{i=1}^{n} x_{i}}
\end{align}
When all quantities are equal, the inequality is minimal and $G=0$, and in the other
extreme case when only one of the quantity is non-zero, then $G=1-1/n$ (which goes to $G=1$ for 
large $n$). More generally, the larger the Gini and the larger the inequality among the variables $x_i$. During each step $t$ of the decimation process we have a set of areas and we can compute the corresponding Gini coefficient (denoted by $G(t)$ or $G(\tau)$).  The algorithm for constructing this `Gini profile' is then the following one:
\begin{enumerate}

\item Associate to each cell $i$ its area $a_{i}$ (if there is a couple $i,j$ such that $a_{i} = a_{j}$, one can infinitesimally perturb $a_{i}$ or $a_{j}$).

\item Find the smallest area size $a_{m}$ and its associate cell $m$.

\item Merge the cell $m$ with its neighbor of largest area size. The new cell has an area equal to the sum of the merged cells, and keeps the same neighbors.

\item Compute the Gini coefficient $G(t)$ for all cells existing at this time $t$.

\item Go back to (2) until there is only one cell left.

\item Plot the Gini profile, \textit{i.e.} the $(t,G(t))_{t \in [0,N-1]}$ graph.
\end{enumerate}

\begin{figure}[!ht]
\begin{center}
\includegraphics[scale=0.8]{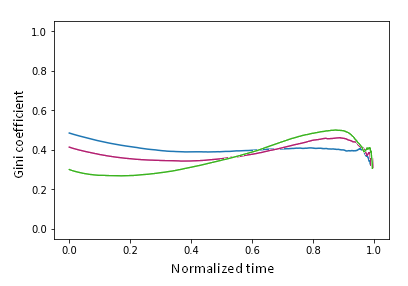}
\end{center}
\caption{Gini profiles versus the normalized time for the Poisson Voronoi graph (in green), the Relative Neighborhood Graph (in red) and the Gabriel Graph (in blue).}
\label{ginicomp}
\end{figure}

\begin{figure}[!ht]
\begin{center}
\includegraphics[scale=0.7]{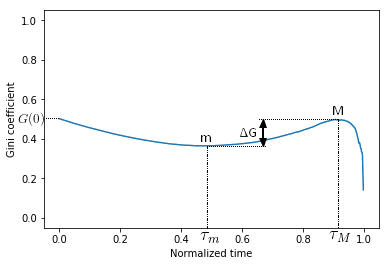}
\end{center}
\caption{Typical profile of the Gini profile. The local minimum $m$ is obtained for $\tau=\tau_m$ and the maximum for $\tau=\tau_M$. The Gini gap is defined as $\Delta G=G_M-G_m$.}
\label{figtypical}
\end{figure}

\begin{figure}[!ht]
\begin{center}
\includegraphics[scale=0.3]{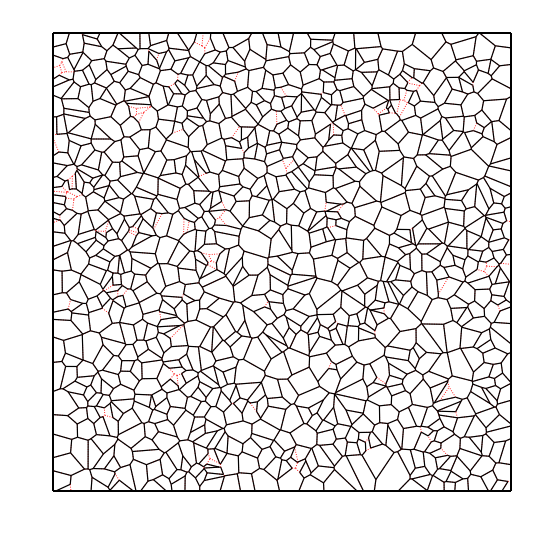}

\includegraphics[scale=0.3]{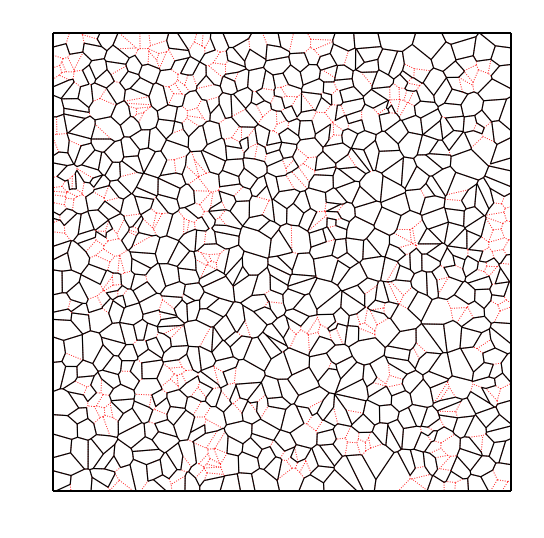}

\includegraphics[scale=0.3]{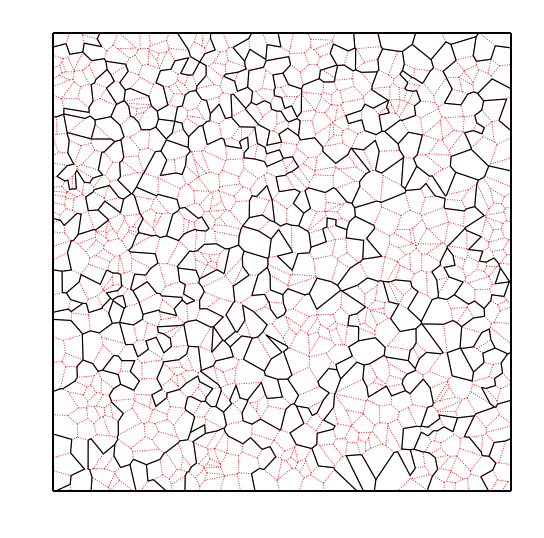}
\end{center}
\caption{Visualization of the graph for three different times of the decimation process applied to the Poisson-Voronoi graph. The red dotted lines represent the deleted edges. Top: graph when the Gini is minimal ($\tau=\tau_m$). Middle: intermediate graph between the minimal Gini and the maximal Gini ($\tau_m<\tau<\tau_M$). Bottom: graph when the Gini is maximal ($\tau=\tau_M$).}
\label{algoex}
\end{figure}

We can then construct a Gini profile for each planar map and we show in Fig.~\ref{ginicomp} the results for the synthetic graphs considered here.
This figure shows that the decimation process produces different responses for these synthetic graphs and seems therefore to be a good candidate for characterizing these networks. From these graphs, we can infer that the typical shape of this Gini profile is as shown in Fig.~\ref{figtypical}.
We represent the corresponding graphs at different stages of the decimation process in the case of the Poisson-Voronoi graph in Fig.~\ref{algoex}.

This typical shape displays first a decrease to a local minimum (at time $\tau=\tau_m$) followed by an increase to a local maximum at $\tau=\tau_M$, and eventually a decrease to the minimum $G=0$ at $\tau=1$. We can understand this typical behavior:
\begin{itemize}
\item The first step (the decrease) is due to the small cells merging with their larger neighbors. When a small cell merges with a large one, the inequality of the area size repartition drops; a comparison would be to merge the bank accounts of two individual, one very rich, and the other with small income.

\item The second step (the increase) is due to the fact that after some time, if all cells are roughly of the same size, merging two cells can only increase inequality, leading to an increase of the Gini coefficient.

\item The third step (the final decrease) is actually a mix of two things. The number of cells is decreasing and at a small scale we observe oscillations until $t=N-1$ (such oscillations can be seen in Fig.~\ref{ginicomp}). Moreover, the Gini coefficient for $t=N-1$ is necessarily equal to $0$, by definition, and this leads on average to a decreasing function.

\end{itemize}

\section{Application to real-world street networks}

We now test the Gini profile on real-world street networks for different cities and understand what information it conveys about the graph. The road networks of cities are obtained from the open database OpenStreetMap \cite{OSM}. We will start with Barcelona (Spain), New York City (USA), Tokyo (Japan) as these large cities combine various neighborhoods with different structures. In a last part we will consider the time evolution of the street network of Paris and see how we can detect structural changes during the evolution of this system.

\subsection{Barcelona, Spain: a first test}

\begin{figure*}[!ht]
\begin{center}
\includegraphics[scale=0.6]{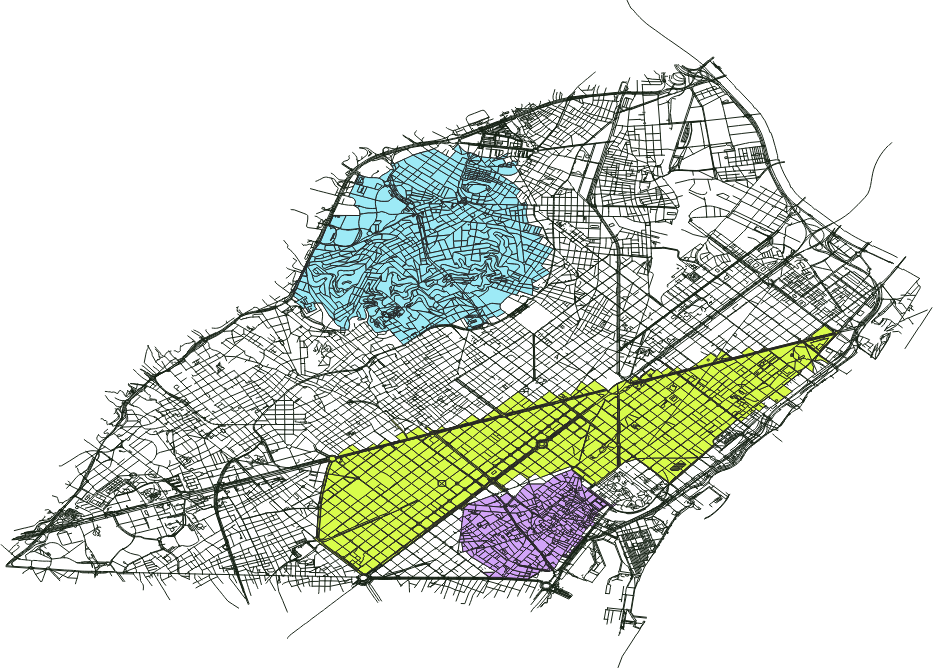}\\
\includegraphics[scale=0.6]{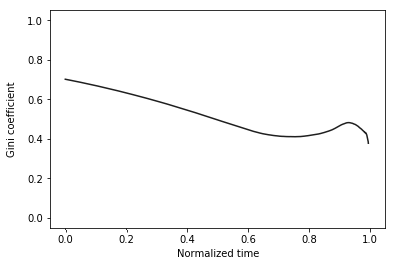}
\includegraphics[scale=0.6]{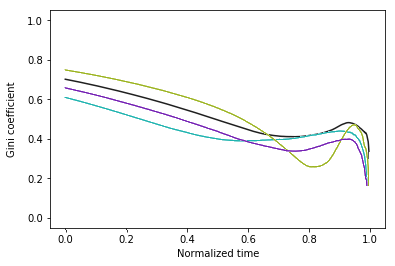}
\end{center}
\caption{(Top) Streets of Barcelona, Spain. Three areas will be studied, referred as blue sector (north), yellow sector (middle) and purple sector (south). (Bottom, left) Gini coefficient over normalized time for the streets of Barcelona. (Bottom. right) Gini coefficient over normalized time for the streets of Barcelona. Each curve represents a different sector of Barcelona, the black curve being for the whole city.}
\label{barcelona}
\end{figure*}

The first city on which the algorithm is tested is Barcelona, Spain. An interesting feature of this city is the presence of many boroughs of different shapes (figure \ref{barcelona}) which gives the opportunity to understand the Gini profile on a practical case.

We first test the algorithm on the whole city. The resulting Gini profile is shown in Fig.~\ref{barcelona}(bottom, left) and corresponds indeed to the typical behavior described above, in particular with the presence of a gap $\Delta G$ (of order $0.1$ here). However this Gini profile alone does not give any information about the components of the city. Indeed, as shown in the Fig.~\ref{barcelona}(bottom, right), the Gini profile displays important variations depending on the borough considered. This last plot shows then that applying the decimation algorithm on a whole city (or more generally to an area that is too large) smoothes out some details that can be important. It also shows that the Gini profile is able to detect variations between different neighborhoods. Finally, this result also suggest that the Gini gap $\Delta G$ is connected to the homogeneity or regularity of the network. For a more homogeneous network, the Gini gap is larger as can be seen for the yellow section of Barcelona and compared to the more random borough in blue or purple (see Fig.~\ref{barcelona}).

\subsection{New York, United States: comparing regular and irregular networks}

\begin{figure}[!ht]
\begin{center}
\includegraphics[scale=0.5]{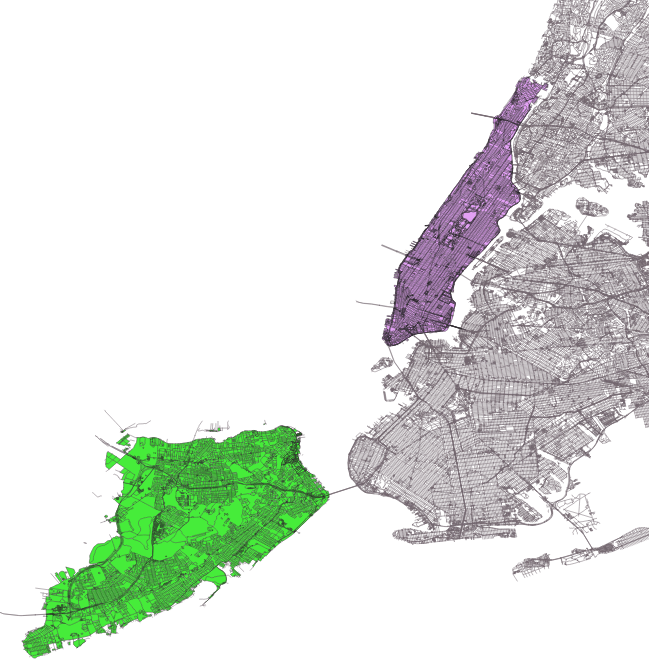}
\end{center}
\caption{Streets of New York, United States. Two areas will be studied: Manhattan (pink) and Staten Island (green).}
\label{newyork}
\end{figure}

\begin{figure}[!ht]
\begin{center}
\includegraphics[scale=0.6]{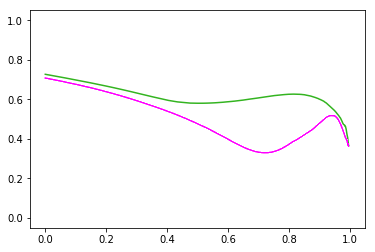}
\end{center}
\caption{Gini curves for Manhattan (in pink) and Staten Island (in green).}
\label{newyorkgini}
\end{figure}

In order to test if there is indeed a link between the Gini gap and the regularity of the street pattern, we now consider another example, the city of New York (USA). This city comprises five boroughs that were developped at different time and have different structures (see for example \cite{Carra:2017} and reference therein). We will focus on two boroughs, Manhattan and Staten Island (see Fig.~\ref{newyork}).
In particular, Manhattan is very regular and comprises essentially rectangular cells \cite{Louf:2014b}, and is subdivided in almost regular patterns with the exception of Central Park in the center of the island. In contrast, Staten Island comprises many green areas and is composed of multiple small neighborhoods at different locations. We computed the Gini profile for these two boroughs and the result is shown in Fig.~\ref{newyorkgini}.
In this figure, we observe that $\Delta G^{\text{Manhattan}} \approx 0.2$ and $\Delta G^{\text{Staten Island}} \approx 0.05$ which confirms the link between regularity of the pattern and magnitude of the Gini gap. We note that this result is also observable in the Fig.~\ref{ginicomp} Gini curves: the gap for the random Voronoi graph (which is statistically homogeneous) is larger than the ones obtained for Gabriel or the Relative Neighborhood graphs.

\subsection{Tokyo (Japan): the meaning of the Gini profile's minimum}

\begin{figure}[!ht]
\begin{center}
\includegraphics[scale=0.5]{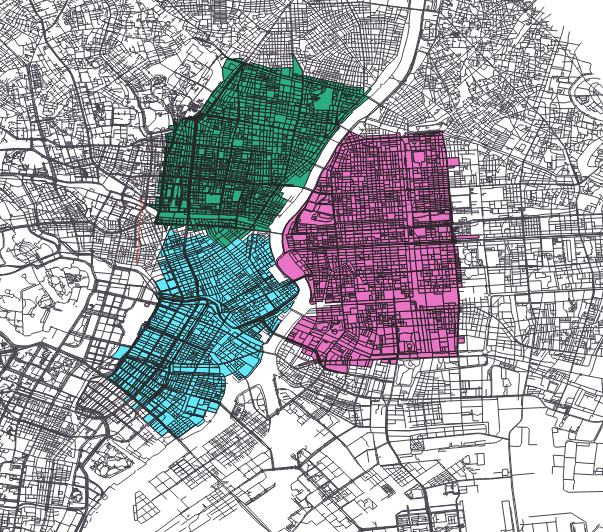}
\end{center}
\caption{Streets of central Tokyo (Japan). We will focus on three different neighborhoods that are represented by different colors: the northwest sector (in green), the southwest sector (in blue) and the east sector (in pink).}
\label{tokyo}
\end{figure}

\begin{figure}[!ht]
\begin{center}
\includegraphics[scale=0.6]{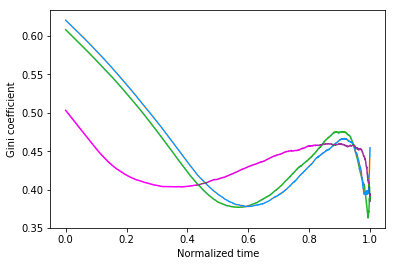}
\end{center}
\caption{Tokyo: Gini profiles for the east sector (in pink), the northwest sector (in green) and the southwest sector (in blue).}
\label{tokyogini}
\end{figure}

\begin{figure}[!ht]
\begin{center}
\includegraphics[scale=0.6]{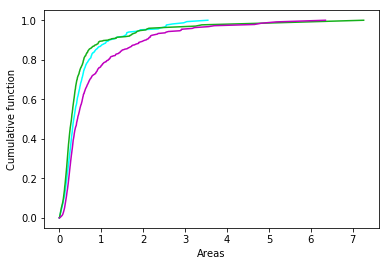}
\end{center}
\caption{Cumulative distributions of the cell's area size for the three sectors of Tokyo considered here (the unit of area size is irrelevant here as we are interested in differences between the sectors only).}
\label{arearep}
\end{figure}

We now consider another large city, Tokyo (Japan) and represent a portion of it in Fig.~\ref{tokyo}. All the different boroughs in the city center are very similar (macroscopically) and we test here if the Gini profile can give a more precise characterization. We thus separate the center in three different sectors as shown in Fig.~\ref{tokyo}, 
and compute the Gini profile for them.
We show these profiles in the Fig.~\ref{tokyogini}.
We observe that the green and blue Gini profiles are very similar, while the pink one (that corresponds to the east sector) is very different, both in terms of the Gini gap but also in terms of the location of both the local minimum ($\tau=\tau_m$) and the maximum ($\tau=\tau_M$) of the profile. A first understanding can be grasped by considering the cumulative distribution of the area size of cells, as shown in Fig.~\ref{arearep}.
We note that the green and blue sectors have similar cumulative functions, while the pink sector displays a broader distribution: there are more cells of very small area in the two first sections that in the pink one. The larger number of small areas will then lead to a larger $\tau_m$ that we indeed observe for the east sector. This is also connected to the initial value of the Gini coefficient that is less important for the pink sector. This shows that there is a larger equality for the block size in the east sector, at odds with the visual impression of Fig.~\ref{tokyo}), probably due to the fact that the pink sector contains apparently more larger blocks. 

\begin{figure}[!ht]
\begin{center}
\includegraphics[scale=0.6]{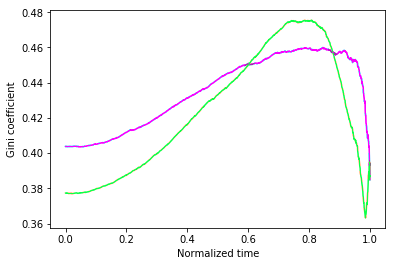}
\end{center}
\caption{Modified Gini curves for the east sector (in pink) and the northwest sector (in green). The blue sector is not plotted, but is similar to the green one, as before.}
\label{renew}
\end{figure}

\section{Paris (France) : a time evolving network}

The city of Paris has been the theather of many changes and the structural evolution of its street network is well documented thanks to the existence of many historical maps (see \cite{Barthelemy:2013} and references therein). Thanks to modern GIS techniques and the digitalization of these historical documents, we are in the position of studying quantitatively the temporal evolution of the street network on almost 300 years \cite{GHD}.

\begin{figure}[!ht]
\begin{center}
\includegraphics[scale=0.6]{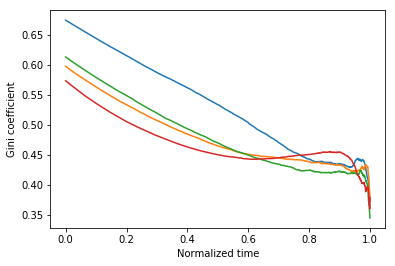}
\includegraphics[scale=0.6]{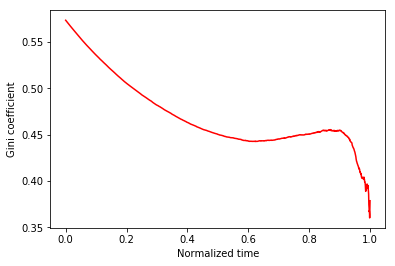}
\end{center}
\caption{(Top) Gini profiles for the central part of Paris, France for the years 1790 (blue), 1836 (green), 1849 (orange) and 1888 (red). (Bottom) The second plot only shows the 1888 Gini curve.}
\label{giniparis}
\end{figure}

One of the most important transformations in Paris was Haussmann's renovation of Paris \cite{Moncan2009}, which took place between 1853 and 1870 under the direction of Napoleon III (we also note that some changes in parts of the city of Barcelona discussed above was actually inspired by Haussmann's work). Haussmann's works had a huge impact on the city of Paris, and maybe surprisingly, none of the standard indicators are able to reveal important changes. It is only by studying the spatial distribution of the betweenness centrality \cite{Barthelemy:2013} that we can observe quantitatively how Haussmann modified in depth the structure of the street network by changing the way we navigate in this system \cite{Barthelemy:2013}. This case is thus a natural playground for testing the relevance of the Gini profile. We have the network for central Paris \cite{GHD} for 4 different dates (1790, 1836, 1849, 1888) among them 3 before Haussmann's works and one posterior to them (we note that although Paris gained ground over time and extended its borders, we only use the algorithm on a central part of Paris corresponding to the year 1790 and which allows us to monitor the changes experienced during this period). The corresponding Gini profiles are shown in Fig.~\ref{giniparis}. 
We observe that there are essentially two main changes highlighted by these Gini profiles. The first one is from year 1790 to 1836 (1836 and 1849 being very similar). This corresponds to the period after the French revolution during which a redistribution of central nodes were observed in this period \cite{Barthelemy:2013}. This is in agreement with the historical fact that after the french revolution many religious and aristocratic domains and properties were sold and divided in order to create new houses and new roads, improving congestion inside Paris. By comparing directly the 1790 and 1836 curves we observe various effects: first the initial Gini coefficient ($G(t=0)$) is smaller, indicating a homogeneization of the size of cells. Second, a local minimum appears, in line with the fact that inequality of cell sizes is smaller in 1836. The last curve of Fig.~\ref{giniparis} for the year 1888 (thus after Haussmann's works) clearly differs from the curves for 1836 or 1849. In particular the initial Gini coefficient is smaller, and maybe more important we observe the appearance of a Gini gap (around $\tau\approx 0.6$). This shows that regularity was introduced by Haussmann in the street network of Paris and that these works led to more regularity in the pattern of blocks. 

Although probably more work is needed here, these results suggest that the Gini profile is indeed a good candidate for characterizing at a coarse-grained level the structure of a planar map. 

\section{Discussion}

A good coarse-grained characterization should smooth out irrelevant details but should keep important 
 structural determinants. The balance between these two antagonistic requirements 
 is the key for a successful characterization of planar maps that can be used for constructing a typology. We proposed here a simple algorithm that maps an embedded planar graph to a curve, the Gini profile.  We showed here that certain features of the Gini profile can be related to the planar map: the position of the local minimum is related to the abundance of small cells, and the Gini gap is related to the statistical homogeneity of the graphs. Also, we showed that various neighborhoods of the same city actually display different Gini profiles, highlighting the sensitivity of this measure. This is confirmed by the study on central Paris where most standard indicators do not display any interesting variations, while the Gini profile can display dramatic variations (before and after Haussmann for example). The Gini profile therefore probably contains some information about spatial correlations, a feature that was absent in the classification proposed in \cite{Louf:2014b}. We believe that this approach could be a first step towards a simple characterization and classification of planar maps, but further tests and studies are certainly needed.

\section{Acknowledgements}

AD thanks the Ecole Normale Sup\'erieure Paris-Saclay for funding
and the IPhT for its hospitality. MB thanks E. Katifori for discussions about this problem and H.S. Dhillon and T. Gross for interesting and stimulating discussions during the conference `Spatially Embedded Networks' that took place in Bristol. 

\section{References}

\bibliographystyle{unsrt}
\bibliography{biblio}

\end{document}